
\magnification=1200
\baselineskip=24pt

\noindent
{\bf INTERSECTION FORMS AND THE
GEOMETRY OF LATTICE CHERN-SIMONS THEORY}
\vskip 0.5truein

\noindent
{\bf D. Eliezer and G. W. Semenoff}\footnote{}{This
work is supported in part by the Natural Sciences and Engineering
Research Council of Canada.}
\vskip 0.1truein

\noindent
{\it Department of Physics, University of British Columbia}

\noindent
{\it Vancouver, British Columbia, Canada V6T 1Z1}
\vskip 0.3truein
\baselineskip=24pt

\centerline{\bf Abstract}
\vskip 0.1truein
We show that it is possible to formulate Abelian Chern-Simons theory
on a lattice as a topological field theory.  We discuss the
relationship between gauge invariance of the Chern-Simons lattice
action and the topological interpretation of the canonical structure.
We show that these theories are exactly solvable and have the same
degrees of freedom as the analogous continuum theories.
\vskip 1.0truein
In the continuum, Chern-Simons theory [1,2] is a topological field
theory [3,4].  It is exactly solvable and is related to interesting
topological structures such as the Witten invariants of
three-manifolds and the Jones polynomial and other knot invariants for
links embedded in three-manifolds.  Its canonical formalism also has
an interesting relationship with conformal field theory in two
dimensions.  As a model for physical phenomena its U(1) (and sometimes
U(N)) versions are proposed as the infrared limit of phenomenological
theories of the fractionally quantized Hall effect[5] and other
theories of anyons, including speculations about the mechanism for
high $T_c$ superconductivity[6].

In recent years there have been efforts to write Chern-Simons theory
on a lattice[7-11].  This gives an automatic ultraviolet
regularization of the field theory and is important for condensed
matter and statistical physics applications.  It also makes
well-defined the correspondence between matter-coupled Chern-Simons
theory and a theory of anyons [11].  Generally, it is difficult to
formulate lattice Chern-Simons theory in a natural way as one lacks
geometrical principles which would replace the general covariance of a
topological field theory in the continuum.

Nevertheless, we shall show in this Letter that when lattice
Chern-Simons theory is formulated so that it is both local and gauge
invariant on the lattice (here the term ``local'' means that the
canonical structure is such that a link commutes nontrivially only
with links with whom it shares a common site), we obtain an exactly
solvable model which shares many of the features of its continuum kin
and can be regarded as a topological field theory on the lattice.

To begin, we review some of the features of continuum U(1)
Chern-Simons theory on the space ${\cal M}^3=R^1\otimes\Sigma^g $,
where $\Sigma^g $ is a compact oriented two-dimensional Riemann
surface with genus $g$.  The action is $$ S~=~\int_{{\cal M}^3}
d^3x\left({k\over4\pi}
\epsilon^{\mu\nu\lambda}A_\mu\partial_\nu
A_\lambda+A_\mu j^\mu\right)
\eqno(1)
$$ The vector density $j^\mu(x)=\int d\tau \sum_i{d\over
d\tau}r_i^\mu(\tau)\delta^3(x-r_i(\tau))$ corresponds to a collection
of Wilson loops parametrized by $r_i^\mu(\tau)$ and $\partial_\mu
j^\mu=0$.  As is usual in gauge theory, the action is linear in the
temporal component of the gauge field and the field equation resulting
from its variation enforces the constraint $$ B(x)+{2\pi\over
k}j^0(x)= 0
\eqno(2)
$$ where $B=dA$ (restricted to $\Sigma^g $) is the magnetic flux.  It is
useful to think of the remaining first order action, $\int
dt\int_{\Sigma^g }\left(-{k\over2\pi}A\wedge{\dot A}-A\cdot j\right)$ as
already being cast in phase space, which is the set of gauge
connection 1-forms on $\Sigma^g $. From this we obtain the canonical
commutator $$
\biggl[ A_i(x), A_j(y)\biggr]~= ~{2\pi i\over k}~\epsilon_{0ij}~\delta^2(x-y)
\eqno(3)
$$
which has the property that, if we consider two distinct
oriented curves, C and C$^\prime$ on $\Sigma^g $,
$$
\biggl[\int_C A,\int_{C'} A\biggr]~= ~{2\pi i\over k}~\nu[C,C']
\eqno(4)
$$ where $\nu[C,C']$ is the number of right-handed minus the number of
left-handed intersections of C and C$^\prime$. (A right-handed
intersection occurs when, if we move along the positive direction of C,
C$^\prime$ crosses from right to left.) Thus, the commutator gives a
representation of the intersection form for Wilson loops.
For closed curves $C$ and $C'$, $\nu[C,C']$ is a topological
invariant and if either $C$ or $C'$ is contractible, $\nu[C,C']$
must vanish.  Therefore, there is a nontrivial commutator only
for homologically nontrivial curves.  If $\alpha_j,
\beta_j$, $j=1,\dots,g$, are a canonical set of closed curves on $\Sigma^g $
generating its first homology group then $\nu[\alpha_i,\alpha_j]=
\nu[\beta_i,\beta_j]=0$, $\nu[\alpha_i,\beta_j]=\delta_{ij}$.
Upon imposition of the constraint (2) the integrals of $A$ over the
1-cycles $\alpha_i,\beta_i$ are the only remaining degrees of freedom
and, modulo the remaining symmetry under large gauge transformations,
they form the reduced phase space[2].

It is the property (4), that the symplectic structure gives the
intersection form for loops on $\Sigma^g$, which we shall discover on
the lattice, as a consequence of lattice gauge invariance and
locality.  We work in continuum time and a square spatial lattice
with spacing 1.  We begin by fixing some notation.  The forward and
backward shift operators are $ S_i f(x) = f(x + \hat i) $ , $ S^{-1}_i
f(x) = f(x-\hat i)~, $ respectively, and forward and backward
difference operators are $ d_i f(x) = f(x+\hat i) - f(x), \ \ d_i =
S_i - 1 $, $\ \
\hat d_i f(x) = f(x) - f(x - \hat i)~, \ \ \hat d_i = 1 - S^{-1}_i =
S_i^{-1} d_i
$.
Summation by parts on a lattice takes the form (neglecting
surface terms)
$
\sum_x f(x) d_i g(x) = - \sum_x \hat d_i f(x) \ g(x)
$ by virtue of the lattice Leibniz rule $ d_i (fg) = fd_i g + d_i f
S_i g = fd_i g + S_i( \hat d_i f g) $ (no sum on $i$).  The components
$A_i(x)$ of the gauge field are real-valued functions on the links
specified by the pair $[x,\hat i]$, $A_0$ is a function on lattice
sites, the magnetic field $ B(x) = d_1 A_2(x) - d_2 A_1(x) $ is a
function on plaquettes where $x$ labels the plaquette with corners
$x,x+\hat1,x+\hat1+\hat2,x+\hat2$, and the electric field
$F_{0i}={\dot A}_i-d_iA_0$ is
a function on links.

A gauge invariant, local, nondegenerate Chern-Simons term was found in
ref. [11] : $$ S =\int dt \sum_{x}\left( {k \over 2 \pi} A_0(x,t)
\epsilon^{ij}d_iA_j(x,t) - {k \over 4 \pi} A_i(x,t)K_{ij}\dot A_j(x,t)+
A_\mu(x,t) j^\mu(x,t)\right)
\eqno(5)
$$ where $i,j=1,2$, with $j_\mu$ a conserved current, $\partial_0
j_0-\hat d_i j_i = 0$\footnote{$^\dagger$}{The divergence of a vector
field is correctly transcribed onto the lattice through backward
differencing.} and $$
\eqalignno{ K_{ij} &= -{1 \over 2}\left( \matrix{ S_2 - S_2^{-1} &
		     -( -1 + S_2^{-1} + S_1 + S_2^{-1}S_1 ) \cr
			-1 + S_1^{-1} + S_2 + S_1^{-1}S_2 &
			 	S_1^{-1} - S_1 \cr } \right)  \cr
		 &=- {1 \over 2}\left( \matrix{ d_2 + \hat d_2 &
		      -2 - 2 d_1 + 2 \hat d_2 + \hat d_2 d_1 \cr
		       2 + 2 d_2 - 2 \hat d_1 - \hat d_1 d_2 &
				-d_1 - \hat d_1 \cr} \right)
&(6)\cr}
$$
Analogous to (3), $K_{ij}^{-1}$ determines the symplectic structure
on the phase space which is the space of functions $A_i(x)$ from the
links of the lattice to $R^1$,
$$
[A_i(x),A_j(y)] = -{2\pi i\over k}
K_{ij}^{-1}(x-y)
\eqno(7)
$$
Also, the equation of motion for $A_0$ gives the gauge constraint
$$
\epsilon^{ij}d_i A_j(x)+{2\pi\over k} j^0(x)=0
\eqno(8)
$$

This associates the magnetic flux in the plaquette with corners
$x,x+\hat 1,x+\hat 1+ \hat 2,x+\hat 2$ with the charge at $x$.  It was
shown in [11] that (7) gives Wilson line-sums the lattice analog of
the algebra (4),
$$
\left[\sum_{[x,i] \in { C}} dx_i A_i(x),\sum_{[y,j] \in { C'}} dy_j
A_j(y) \right]  ={2\pi i\over k} \nu({ C},{ C}')
\eqno(9)
$$
where the sums are over lattice curves (connected assemblies of
oriented links) and $dx_i=\pm1$ according to whether the link $[x,i]$
is backward or forward directed.

We may work backwards from (9), and begin by considering it a
primitive requirement that $K^{-1}$ is anti-Hermitean, nondegenerate
and gives the symplectic structure in (9).  In particular, if $C$ in
(9) is a homologically trivial closed loop and $C'$ is an open curve,
then $\nu[C,C']$ is an integer, the number of times $C$ links each
endpoint of $C'$, with sign determined by the orientations of $C$ and
$C'$.  This means that $K^{-1}$ must obey
$$
\hat d_i =d_j\epsilon_{jk}K^{-1}_{ki}~~,
{d}_i=-K^{-1}_{ij}\epsilon_{jk}{\hat d}_k~~,
\eqno(10)
$$
This turns out to be identical to the condition on $K$ that one
obtains by requiring that the action (5) is invariant under the gauge
transform $A_i\rightarrow A_i+d_i\chi$, $A_0\rightarrow A_0+\dot\chi$,
or equivalently that ${k\over2\pi}B(x)$ is the generator of static
gauge transforms,
$$
{k\over2\pi}
\biggl[\sum_x\chi(x)B(x),A_i(y)\biggr]=-id_i\chi(y)
\eqno(11)
$$ Thus, the local form of the gauge constraint in (8) as well as
gauge invariance of the action are equivalent conditions to the
topological invariance and integer-valuedness of $\nu[C,C']$ in (9)
when either $C$ or $C'$ is a closed contour which is the boundary of a
set of plaquettes.    Pictorially, gauge invariance, together with the
locality of the symplectic structure $K^{-1}$ guarantee that curves
which touch but do not penetrate are counted as zero intersection:
\vskip 1.25truein
Conversely, these
conditions, together with the requirement that $K^{-1}$ allows interactions
of links with only those other links that share a common site, and
that it have a local inverse, may be summarized by precisely the condition (10)
(which we found by requiring gauge invariance), and makes $K^{-1}$ the kernel
that counts
intersections of two lattice curves.
The remaining structure of $K^{-1}$ is the normalization, and is fixed
by requiring that (9) should hold for the simple crossings
\vskip 1.0truein
\vfill\eject
This, together with the Gauss Law\footnote{$^\dagger$}{In fact, there
are exactly four such $K$'s, corresponding to the four possible ways
of assigning a plaquette to one of its corners.  Gauss' Law in these
cases would appear as e.g. $B + {2 \pi \over k}S_1 j_0 = 0$.  }
determines $K^{-1}$ and $K$ uniquely as that given in equation (6).
Several results follow for partial intersection numbers:
\vskip 3.0truein
These would be ambiguous in the continuum and on the lattice they
follow uniquely from our other requirements on $K$.

Note that the Kernel in the action, $S_{CS} = \int dt \sum_x A_\mu(x)
D_{\mu \nu}(x-y)
A_\nu(y)$ here can be parametrized as $D_{\mu \nu} =
T_{\mu \lambda} \epsilon_{\lambda\sigma\nu} d_\sigma$ and
that the requirement of gauge invariance is precisely
$\hat d T = d$ (where $d_0=\hat d_0=\partial_0$), the analog of (10).  When
we require that $T$ is a local matrix, with a local inverse, (the
determinant is a monomial in shifts) this
$T^{-1}$-matrix is the kernel which counts the
intersections of surfaces
with curves on a 3-dimensional lattice. This, in turn implies that
(as we shall see in the following) the effective action for Chern-Simons
theory coupled to currents is related to the linking numbers of trajectories.

To solve the model (3), we first consider a lattice with trivial first
homology group, such as the latticized open plane $R^2$.  We represent
the commutator (5) by choosing the functional variables $$ B(x)={2\pi
i\over k}{\partial\over\partial\lambda(x)}~~,~~\lambda(x)={1\over
d\cdot\hat d}\left(\hat d\cdot A-{\hat d K^{-1}d\over2d\cdot\hat d}B(x)\right)
\eqno(12)
$$
Then, a wave-function which solves the constraint (8) is
$$
\Psi_{\rm phys}[\lambda,j_\mu,t]=\exp\left(i\sum_x
\lambda(x)j_0(x,t)\right)~\tilde\Psi[j_\mu,t]
\eqno(13)
$$ The Hamiltonian is
$$ H=\sum_x j_i(x)A_i(x)=\sum_x\left(j\times\hat d
{1\over d\cdot\hat d}B(x)+ j\cdot d{1\over d\cdot\hat d}\hat
d\cdot A\right)
\eqno(14)
$$
where we have used the identity $\delta_{ij}=\epsilon_{ik}\hat
d_k{1\over d\cdot\hat d}\epsilon_{jl}d_l+d_i{1\over d\cdot\hat d}\hat
d_j$.  The Schr\"odinger equation
$$i{\partial\over\partial t}\Psi_{\rm phys}[\lambda,j_\mu,t]=H\Psi_{\rm
phys}[\lambda,j_\mu,t]
\eqno(15)
$$
is solved by
$$
\Psi_{\rm phys}=\exp\left(i\sum_x\lambda(x) j_0(x,t)-{2\pi i\over k}
\int_{-\infty}^t\sum\left(j\times\hat d{1\over d\cdot\hat d}j^0+ j\cdot
d{1\over d\cdot\hat d}{\hat d K^{-1}d\over2 d\cdot\hat d}j_0\right)\right)
\eqno(16)
$$

The trajectory of a particle is a piecewise linear lattice curve
consisting of instantaneous hoppings in spatial directions between
lattice sites and temporal segments representing the particle at rest
on a particular site.  The second term in the phase of the
wave-function in (16) is a topological invariant and can be
interpreted as the linking number of periodic lattice trajectories.
To see this, first consider adding a closed spacelike curve to
$j^\mu$, the perimeter of a plaquette.  This is described by the
change in current $$
\delta j^0(x,t)=0
{}~~,~~
\delta j^1(x,t)=-\hat d_2\delta(x-a)\delta(t)
{}~~,~~
\delta j^2(x,t)= \hat d_1\delta(x-a)\delta(t)
$$ which has the property that $\hat d\times\delta j(x,t)=d\cdot\hat
d\delta(x-a)\delta(t)$, implying that the phase of the wavefunction changes by
${2\pi\over k}\theta(t)j^0(a,0)$, which is $2\pi\over k$ times the
linking number of the lattice curves with the plaquette.  If we add a
time-like plaquette, for example $$
\delta j^0(x,t)=\theta(t-t_1)\theta(t_2-t)\hat d_1\delta(x-a)
{}~~,~~
\delta j^1(x,t)=(\delta(t-t_1)-\delta(t-t_2))\delta(x-a)
$$ The phase changes by $$ {2\pi\over k}\int_{t_1}^{t_2}dt j^2(a,t) $$
which is also $2\pi/k$ times an integer, the number of times the
lattice curves link the time-like plaquette.  To get the latter result
we have neglected the `self-linking' of the space-like plaquette with
itself.  Such self-linking numbers are not well-defined here but
require further regularization. (Here, self-linking numbers involve
ill-defined products of the form $\theta(t)\delta(t)$.)

In general, a lattice current due to a charged particle which hops
from lattice site $x_{p-1}$ to $x_p$ along links $C_p$ at time $t_p$
has the form $$
\eqalignno{j^0(x,t)&=\sum_p \delta(x-x_p)\theta(t_{p+1}-t)\theta(t-t_p)\cr
j^i(x,t)&=\sum_p\sum_{[y_p,i_p]={\rm  links ~in} ~
C_p}\delta(x-y_p)dy_{i_p}\delta(t-t_p) &(17) \cr}
$$
A general conserved current may be written as a sum of currents of the
form (17) $j_\mu = \sum_\alpha j_\mu^\alpha$.  The quadratic self-interaction
terms for currents in (16) are
of necessity ambiguous, because they are in fact the self-linking
number of the trajectories -- the latter is a topological invariant of {\it
framed} links, and the ambiguity can only be lifted by introducing a
framing.

The cross terms of (16) between two trajectories $\alpha,\ \beta$ are
also ambiguous unless the instantaneous hoppings in spatial directions
of the two lattice curves occur at distinct times.  When this
condition is met, the cross term is well defined.  The $\alpha,\beta$
cross term may be combined with the $\beta,\alpha$ cross term, and
calculated in terms of a lattice angle function.  Lattice angle
functions have been discussed in earlier works [8,11], and are usually
defined as a contour sum as $\theta_{\cal C}(x,y) = 2\pi\sum_{\cal C}^x d
\ell_i{\hat d_i^\perp \over d \cdot \hat d} \delta(\ell - y)$
and are multivalued functions of position in that they depend on
windings of the contour ${\cal C}$ around the point $y$.
$$
\eqalignno{ d\times d \theta_C(x,y) &= 2 \pi \delta(x-y) &(18a) \cr
\theta_{C}(x,y) - \theta_{C'}(x,y)& = 2\pi\sum_z \omega(CC'^{-1},z)
&(18b) \cr}
$$
where $\omega(CC'^{-1},z)$ is the winding number of the
closed curve $CC'^{-1}$ around the point $z$.  Unlike the continuum
angle function, which satisifies the further identity
$$
\theta_C(x,y) - \theta_{C'}(y,x) = \pi +2\pi\nu[C,C']
\eqno(19a)
$$
this angle-function has the property that
$$
\theta_C(x,y) - \theta_{C'}(y,x) = \pi +2\pi\nu[C,C']+\xi(x-y)
\eqno(19b)
$$
The last term in (16) is precisely the defect function
$\xi(x-y)$ for the lattice angle function which the first term in (16)
calculates.  Thus the result for the cross terms $\alpha,\beta$ and
$\beta, \alpha$ together is proportional to
the total change in angle between the curves accumulated during
the time evolution,
$$
{1\over k}\int dt~{d\over dt}{\tilde\theta}(x(t)-x'(t))
{}~~,~~
x\in\alpha~~x'\in\beta
$$
(Note that the result is independent of $C$.)  where $\tilde \theta$
is an ``improved'' lattice angle function, in that it satisfies (19a)
as well as (18ab).  Thus for a periodic trajectory of $N$ particles on
the lattice the phase of the wave-function is the linking number of
trajectories and the wave-function therefore carries a 1-dimensional
unitary representation of the $N^{\rm th}$-order braid group of the
plane where braiding is constrained to follow links of the lattice.
This implies that particles coupled to the lattice Chern-Simons theory
are anyons with statistics parameter $1/k$.  Furthermore, aside from
this phase the theory is trivial.  The Hilbert space contains only one
state.

A more complicated situation arises when the lattice has a nontrivial
first homology group. Here, for simplicity we shall set $j_\mu=0$ and
consider the example of a toroidal lattice where the gauge fields have
periodic boundary conditions
$$
A_i(x_1+N_1,x_2)=A_i(x_1,x_2)~~,~~A_i(x_1,x_2+N_2)=A_i(x_1,x_2)
\eqno(19)
$$
Here, the gauge group is U(1) and we require that the gauge
transformation obeys the boundary condition
$$
\chi(x_1+N_1,x_2)=\chi(x_1,x_2)+2\pi
m_1~~,~~\chi(x_1,x_2+N_2)=\chi(x_1,x_2)+2\pi m_2
\eqno(20)
$$ where $m_i$ are
integers.  We can choose the canonical generators of the first
homology and the gauge invariant canonical variables as
$$ q=\sum_n
A_1(n,0)~~,~~p=-{k\over2\pi}\sum_n A_2(0,n)
\eqno(21)
$$
which obey the algebra
$$
\bigl[ q,p\bigr]=i
\eqno(22)
$$
The large gauge transforms $q\rightarrow q+2\pi
m_1$, $p\rightarrow p-km_2$ are generated by $m_1$ and $m_2$
operations of the unitary operators $g_{1}= e^{2\pi i p}$
$g_{2}=e^{ikq}$, respectively. These operators obey the algebra
$$
g_1g_2=g_2g_1 e^{2\pi i k}
\eqno(23)
$$
The wave-function should carry a
unitary represention of this algebra.  Here we are assuming that $k=k_1/k_2$
is a rational number.  	It is straightforward to construct a
representation of (23).  Assume that we find an eigenvector
$\psi_\theta$ of $g_1$ such
that
$$
g_1 \psi_\theta=e^{i\theta}\psi_\theta~.
\eqno(24)
$$ Operating $g_2$ $\ell$
times gives another eigenvector of $g_1$ with eigenvalue
$e^{i(\theta+2\pi k\ell)}$,
$$
g_1~ g_2^\ell \psi_\theta~=~ g_2^\ell g_1 e^{2\pi ik\ell}\psi_\theta ~
=~e^{\theta+2\pi ik\ell}~g_2^\ell\psi_\theta
\eqno(25)
$$
When $\ell=k_2$ we obtain the original eigenvalue.  Thus
we conclude that the algebra (23) may be represented by $k_2\times k_2$ unitary
matrices and that $g_2^{k_2}\psi_{\theta\xi}=e^{i\xi}\psi_{\theta\xi}$.
Thus, the representation is specified by two angles $\theta$ and
$\xi$.  In the Schr\"odinger polarization the wavefunction is a
function of $q$ and (24) implies
$$
\psi_{\theta\xi}(q)=\sum_{n\in{\cal Z}}e^{(n+\theta/2\pi )iq}\psi_{\theta\xi}
(n)
\eqno(26)
$$
and
$$
g_2^\ell\psi_{\theta\xi}(q)=\sum_{n\in{\cal
Z}}e^{(n+\theta/2\pi+k\ell)iq}\psi_{\theta\xi}(n)
\eqno(27)
$$
When $\ell=k_2$ we obtain
$\psi_{\theta\xi}(n)= e^{i\xi}\psi_{\theta\xi}(n+k_1)$.  Thus, for each
$\ell$ there are $k_1$ independent coefficients in the expansion (26).
The Hilbert space is $k_1k_2$-dimensional.  This construction can be
generalized to lattices with more complicated homology.  For a lattice
with genus $g$, the dimension of the Hilbert space turns out to be
$(k_1k_2)^g$.  This dimension is the same as that of the Hilbert space in
the continuum theory on the space $R^1\times\Sigma^g $ where $\Sigma^g $ is
a Riemann surface of genus $g$.

In conclusion, we have shown that the present version of lattice
Chern-Simons theory, where the form of the action is deduced from the
requirements of locality of the action and canonical structure, as
well as from gauge invariance, recovers the topological features of
continuum Chern-Simons theory.  An interesting feature is the
introduction of a new and practically unique lattice kernel which can
be regarded as an intersection density and which is responsible for
the topological nature of the theory.  We showed that the lattice
model is exactly solvable.  The solution resembles the solution of the
analogous continuum Chern-Simons theory.

The wave-functions carry a representation of the Braid group.
Note that, since particles are constrained to occupy lattice sites,
their motion is more restricted than in the continuum, and it is
possible to form clusters of particles such that not every braiding
operation is allowed.  Thus, the braid group on the lattice
must differ from the one in the continuum by
certain constraints. It would be interesting to investigate this
further.

It is intriguing to us that the Hilbert space of lattice Chern-Simons
theory on the torus is identical to that in the continuum.  In
previous work on the continuum theory [2] the Hilbert space was
related to quantization of the moduli space of the torus, i.e. the
space of metrics on the torus modulo diffeomorphisms.  On the lattice
there are no diffeomorphisms or metrics and no concept of moduli
space. Yet we find a quantum theory on the lattice virtually identical
to that in the continuum.

\vfill\eject
\centerline{\bf References}
\vskip 0.5truein
\item{1.}A.S.Schwarz, Lett. Math. Phys. 2(1978), 247;
Comm. Math. Phys. 67 (1979), 1.
\item{2.}E.Witten, Comm. Math. Phys. 121 (1989), 351.
\item{3.}E.Witten, Comm. Math. Phys. 117 (1988), 353.
\item{4.}D.Birmingham, M.Blau, M.Rakowski and G.Thompson, `Topological
Field Theory', to be published in Physics Reports, 1992.
\item{5.}T.H.Hansson, S.Kivelson and S.-G.Zhang, Phys.
Rev.Lett. 62 (1989), 82;
G. Moore and N. Read, Nucl. Phys. B360 (1991), 362.
\item{6.}A. Fetter, R. Hanna and R. Laughlin, Phys. Rev. B40, 8745 (1989).
\item{7.}J. Fr\"ohlich and P. Marchetti, Comm. Math. Phys. 116 (1988),
127.
\item{8.}M. L\"uscher, Nucl. Phys. B326 (1989), 557.
\item{9.}V.F. M\"uller,  Z.Phys.C47:301-310,1990.
\item{10.}D.Eliezer, G.W.Semenoff and S.S.C.Wu, Mod.  Phys. Lett. A, in press,
1991.
\item{11.}D.Eliezer and G.W.Semenoff, Ann. Phys. (N.Y.), in press, 1992.
\end
\item{12.}R. Sorkin, J. Math. Phys. 16 (1975), 2432
\item{13.}T. Honan, Doctoral Thesis, Univ. Maryland, 1985, unpublished
\item{14.}H. Becher and P. Joos, Z. Phys. C15, 343 (1982).
\item{15.}D. Eliezer and G.W.Semenoff, Phys. Lett. B266 (1991), 375; R. Kantor
and L. Susskind, `A Lattice Model of Fractional
Statistics', Nucl. Phys. B, 1990.

The interesting feature of this result is the matrix $K_{ij}$, which
is responsible for these topological properties.  We devote the rest
of this paper to understanding this matrix and its origins.  To do so,
we latticize Euclidean time, and recast (9) as $$ S_{CS} =
\sum_x\left({ik \over 4
\pi} A_i D_{ij}A_j+A_\mu j^\mu\right)
\eqno(15)
$$ with $D_{\muj }= T_{\mu\lambda}
\epsilon_{\lambda\sigma\nu}d_\sigma$, and $$ T_{\mu\lambda} = \left(
\matrix{S_0&0\cr 0& K_{ij} \epsilon_{jk}\cr
} \right) ~~,~~ T^{-1}_{\mu\lambda}=\left(\matrix{ S_0^{-1}&0\cr
0&-\epsilon_{ij}K^{-1}_{kj}
\cr}\right)
\eqno(16)
$$ Note that $T^{-1}$ is a local matrix.  If we recall the property
(11) of $K_{ij}$ which ensured gauge invariance for (9), we may deduce
that $T$ satisfies $
\hat d T = d$.

The matrix $T$ is also an intersection density in 3-dimensions.
Consider an oriented surface ${\cal S}$ (a collection of oriented
plaquettes) and the antisymmetric field $S_{\mu\nu}(x)=+1(-1)$ if
$x,x+\mu,x+\nu,x+\mu+\nu$ is a positively(negatively) oriented
plaquette in ${\cal S}$ and $S_{\mu\nu}=0$ otherwise.  The surface can
also be described by its normal vector field
$N_\mu(x)={1\over2}\epsilon_{\mu\nu\lambda}S_{\nu\lambda}(x)$.  Then
${\hat d}_\mu S_{\mu\nu}=B_\nu$ is the current density with support on
the boundary $\delta{\cal S}$ and ${\hat d}_\mu B_\mu=0$.  If ${\cal
S}$ is a closed surface, $B_\mu=0$.  We also consider an oriented
lattice contour $C$ with endpoints at $x_f$ and $x_i$ sepcified by the
current density $j_\mu(x)=+1(-1)$ if $[x,x+\mu]$ is a
positively(negatively) oriented link in $C$ and ${\hat d}_\mu
j_\mu(x)=\delta_{x,x_f}-\delta_{x,x_i}$.  If $C$ intersects only
interior points of $S$ then $$
\sum_x N_\mu(x)T_{\mu\nu}j_\nu(x)= \nu[{\cal S},C]
\eqno(17)
$$ where $\nu[{\cal S},C]=\sum_{\rm int.}\pm1$ is the signed
intersection number of ${\cal S}$ and $C$.  (+1(-1) occurs when $C$
and $N$ are parallel(anti-parallel).)  This can be seen by considering
${\cal S}$ a single plaquette and $C$ a single link.  Then if $C$ is
temporal,
\vskip 0.5truein
and all others vanish.  If $C$ is spatial,
\vskip 1.0truein
and all others vanish. If $C$ only intersects points or else links in
the interior of ${\cal S}$ the result (17) follows.  Also, situations
which would be ambiguous in the continuum here are
\vskip 1.0truein

On latticized $R^3$ we can compute the quantum partition function for
the field theory with action (15), $Z = \exp iI$, with $$ I = {i
 \pi \over k}i
\sum_{x,y} \sum_{a,b} j^a_\mu(x) D_{\mu \nu}^{-1}(x-y) j^b_\nu(y)
\eqno(18)
$$
Using the above parametrization of $D$ in terms of $T$, we may write
$D_{\mu \nu}^{-1}$ as\footnote{$^\dagger$}{$D$ is actually noninvertible
because it has a kernel $\hat d D = D d = 0$ -- we have inverted $D$
in its nonzero subspace, using the left inverse, so that $D^{-1}D = P
= I - {d \otimes \hat d \over d \cdot \hat d}$.
This does not cause a problem in the above
calculation because the current $j$ is conserved, and is therefore
orthogonal to the kernel of $D$.}
$$
D_{\mu \nu}^{-1} =
{ \epsilon_{\mu \sigma \lambda} \hat d_\lambda \over d \cdot \hat d}
T_{\nu \sigma}^{-1}
\eqno(18)
$$ and, because $j^a$ is a homologically trivial closed curve, we may
write it as $j^a_\mu = \hat d_\nu S^a_{\nu \mu} = \epsilon_{\mu
\lambda \sigma}
\hat d_\lambda N^a_\sigma$.
Thus, we find
$$
I =
{2 \pi\over k}\sum_{a<b}
\sum_{x,y}
N^a_\sigma(x)
T_{\sigma \nu}^{-1}(x-y)
j^b_\nu(y)
+{\pi\over k}\sum_aN^a_\mu(x)T^{-1}_{\mu\nu}(x-y)j^a_\nu(y)
\eqno(19)
$$
We argue now that $T^{-1}$ precisely counts intersections between a
lattice surface and a lattice contour.  To do so, we first argue that
an intersection form of this kind must be local, in the sense that it
must give zero for a link and plaquette that do not share a common
site.  This is clear, because such a configuration may appear as part
of a larger one in which surface does not intersect with a link at
all.  We then note that an intersection form must also give zero on a
variety of situations where links and surfaces do share a common site
or link, but in which the contour does not penetrate.  An exhaustive
list of such situations (up to permutations of the directions $ ~\hat
0, \hat 1,
{}~\hat 2$ and linear dependency) are shown in Fig. 1.  These
requirements may be recast, however, as the requirement that $T^{-1}$
vanish when summed over any closed surface, and with a contour that
has no boundary in the interior of the lattice.  This can be seen by
adding a small bubble $a'$ to the surface $a$ in expression (26) (see
Fig. 2), noting that this cannot alter the intersection number, and in
fact corresponds to the difference between one of the Figs. 1 and a
nonlocal situation which vanishes by our locality argument.  Call the
additional contribution to (26) $\delta I$. A contour
with no interior endpoints obeys the current conservation law $\hat d
\cdot j = 0$, and a closed surface obeys $\hat d
\times a' = 0$.  To show that $T$ in fact does satisfy these
requirements, we potentiate $a'$ as $a' = \hat d \phi$, with $\phi$ some
local scalar function on the lattice (i.e. a function with finite
support -- this may always be done).
$$
\delta I =
2 \pi \alpha
\sum_{x,y} \sum_{a,b}
a^{\prime a}_\sigma(x)
T_{\sigma \nu}^{-1}(x-y)
j^b_\nu(y)
\eqno(27)
$$
{}From there we may sum by parts,
use property (20) for $T$ and current conservation to show in fact that
$\delta I$ vanishes.

	It remains only to show that $T^{-1}$ counts $+1$ for
penetrations in the positive directions.  This may be done by direct
calculation in the latticized directions using expressions (14) and
(19), and, in the case of continuous time, by simply noting the timelike
delta-function in $T$, and that it is diagonal in the time direction.

The remaining terms in the sum (26), those for which the curve indices
$a$ and $b$ are equal, also represent topological invariants of the
curve.  It is known as the self-linking number, and in the continuum
requires a regularization for a well-defined computation.  This is
because the self-linking invariant is actually an invariant of framed
links, and the regularization supplies this framing.  It is a somewhat
unappealing feature of Chern-Simons theory in the continuum that this
framing does not come from the definition of the theory, but rather
appears as an ambiguity that must be fixed in computations.

Conversely, it is an appealing feature of our lattice definition of
the theory that the expression for the self-linking number is
nonsingular and well-defined.  This means that the framing is not an
ambiguity of the theory, but has somehow been implicitly specified by
our choice of lattice conventions.  The choice made is simply that
framing which gives zero self-linking for a plane curve, with all
other curves being obtained from it by a sequence of moves consisting
of adding or subtracting plaquettes from it, distorting it, or moving
it through itself.  All moves but the last do not affect the
self-linking number, and the last increases it by two (this is easier
to see in the continuum because, when the framing is explicit, there
are two intersections of the same sign made between the curve and it's
parallel partner).  A result of this definition is that the
self-linking number in these topological field theories counts only
the number of intersecting moves made, and the twist and writhe changing
precisely to mirror this result.

To see that this lattice self-linking number is in fact a topological
invariant, it is sufficient to add an additional plaquette to the
curve and surface in quantity (26) (i.e. one of the terms with $a=b$),
and show that the additional contribution to (26) vanishes.  Such a
contribution would have three terms (See Fig. 2a).  The last would
simply be the self-linking of the plaquette itself, which it may be
checked vanishes for $T$ given by (14) and (19), and actually may be seen to
follow trivially if symmetry among the 3 directions is imposed, though
does not directly seem to follow from (20), and may thus be
thought as as an additional requirement on an intersection matrix.
The other two terms may be recombined (see fig. 2b), together with the
conditions represented in Figs. 1, (equivalent to (20)),
into another of the same conditions.

	Note that little more was required of $T$ than property (20),
the only additional requirements being the vanishing of the
self-linking number of the curve bounding a single plaquette, and that
it not vanish on penetrations.  Indeed, rather than (20), one needs
only the weaker condition $\hat d T \propto d$.  Thus one sees
immediately that the requirements of gauge invariance are contained in
those for intersections.  But we also see that these requirements form
a linear system of equations which vastly underdetermines $T$, and
thus there are a large family of these $T$ matrices, which we call
``intersection forms'', because they may be integrated up over
homology classes to obtain the intersection form used in homology
theory.  We may thus contemplate imposing other conditions.  Adding
symmetry conditions fixes little of this freedom, as many of the
conditions we impose are actually eliminated by it.  A desiderata
would be a local inverse (i.e. the determinant is a monomial in
shifts), because this would allow us to write down other local lattice
Chern-Simons actions.  Equation counting implies that this condition
vastly overdetermines the system -- nevertheless, there are several
solutions in addition to (19).  One of these has already been found
earlier in the context of a geometric description for lattice
fermions[14], and then later discussed for writing down a geometric
description of a lattice Chern-Simons term for a Maxwell-Chern-Simons
theory[7,8,9], though this effort could not be carried over directly
to topological field theories, or to free anyons (i.e. not strongly
coupled to a photon), because, when the Chern-Simons term contains the
canonical structure, the $K$ matrix in equations (14) and (19) is
required to be anti-hermitian in addition [see 11].  However, this
restriction is not necessary in more general cases, such as when two
independent vector fields $B, A$ are coupled by a Chern-Simons like
term of the form (in general dimension) $B \hat dA$.  The intersection
form found by Becher and Joos was the simplest one
$$
T =		\left(	\matrix{S_0 & 0 & 0 \cr
				0 & S_1 & 0 \cr
				0 & 0 & S_2 \cr} \right)
\eqno(28)
$$
Another, nonsymmetric, one suitable for taking the time continuum limit is
$$
T = {1 \over 2}\left( \matrix{  -( -1 + S_2^{-1} + S_1 + S_2^{-1}S_1 ) &
		     S_1 - S_1^{-1} & 0\cr
			 S_2^{-1} - S_2 &
			 -1 + S_1^{-1} + S_2 + S_1^{-1}S_2 & 0 \cr
			0 & 0 & 2 S_1^{-1} S_3 \cr } \right)
\eqno(29)
$$
There is another symmetric one as well, but it is too long to include
in the text.

While it seems remarkable that the requirements on the lattice
coincide so nicely, there is in fact a simple reason for this.  The
requirement (20) on $T$ which allows summation by parts to proceed as
in the proof of gauge invariance of (13) is a kind of lattice Leibniz
rule, which permits one to define a lattice algebra analogous to the
exterior algebra of differential forms.  For example, we may
immediately write the action (13) for Hamiltonian lattice Chern-Simons
theory in terms of forms defined on a 2-dimensional spatial lattice,
and letting the matrix $K$ define the wedge product.  The algebra
obtained thereby satisfies a Leibniz rule, Stokes' theorem, and the
anti-symmetry of a normal exterior differential system.  A higher
dimensional generalization may be used to write the Chern-Simons
action (18) in terms of lattice forms, but we must define the wedge
product of 1-forms with 1-forms differently from, say 1-forms with
2-forms, and the result is a non-associative algebra of lattice forms.
This geometric interpretation is therefore not a straightforward
transcription of the machinery used in the continuum.  However, a
sensible cohomology may be defined in this formalism, and it in fact
provides a very direct correspondence with cubical homology, i.e. that
a cycle $j_\nu$ surrounding a hole, a homology generator, may be seen
also as a closed, nonexact form, i.e.  a cohomology generator, using
the relations given in the footnote after equation (21).  (This in
fact may be generalized to arbitrary dimension.)  Thus, given the
close links outlined above to gauge invariance, topological field
theories, and a geometrical theory in 2 dimensions, we are confident
of the relevance of this geometrical interpretation.  Further, in a
different context, a lattice exterior system was actually erected by
Becher and Joos[14]using the $T$-matrix in (28), though the connection
with intersections on the lattice was unknown.

Finally, we note that none of these considerations need be confined to
$2+1$ dimensions.  The existence of intersection forms in any
dimension is established by the equation counting arguments mentioned
previously, and indeed at least the Becher-Joos formula generalizes
trivially.  In particular, the three-dimensional intersection forms
$T$ may be used to latticize the $3+1$ dimensional topological field
theory (5), by giving the commutator the topological property of
intersections, allowing for a higher dimensional version of the
commutators (4).  Thus in general dimension the above discussion gives
a prescription for latticizing any topological field theory -- one
simply finds an intersection form to insert into the naive
latticization of the theory, which then plays the role of the
canonical structure.  The only caveat is that which occurs in
Chern-Simons theory, namely that if the field variables occur so that
under summations or integrations by parts one gets back the same term,
it may impose additional symmetry requirements (e.g. anti-hermicity)
on the canonical structure which may spoil the intersection property.

In this letter, we have discussed the connections between gauge
invariance and topological properties of a lattice field theory.  We
found that, in any dimension, it is possible to latticize a
topological field theory, by inserting a matrix which has the
interpretation of an intersection form density.  This means that, when
summed over arbitrary $p$- and $d-p$-forms and modded out by homology
class, the usual intersection form is obtained.  Such matrices exist
on square lattices in any dimension, and were shown to lead to gauge
invariance always.  In fact, they also give a well-defined
prescription for a lattice exterior algebra, whose existence may be
seen to be the root of all of these connections.

The intersection-form density, $K^{-1}_{ij}$, can be used to construct
an exterior product for differential forms on the lattice. On a
lattice, a scalar field lives on sites, a vector field on links, a
tensor field on plaquettes, etc.  The difference-form of a scalar function
$ d\Lambda(x)\equiv dx^i d_i\Lambda(x) $ has the property that under a
line-sum where the line has endpoints $x$ and $y$, $\sum d\Lambda=
\Lambda(y)-\Lambda(x)$.  Furthermore, if we consider a vector field,
$A_i(x)$ for example, and we make the 1-form $A_idx^i$.  To make a
2-form we either take the differential of a 1-form,
$dA=d_iA_jdx^i\wedge dx^j$ or the product of two one-forms, $AB=A_iB_j
dx^i\wedge dx^j$.

The existence of such an algebra actually guarantees both existence of
gauge invariant actions, and the existence of intersection forms.  We
may use the 2 and 3 dimensional intersection forms to define wedge
products which will, with the natural definition of exterior
differencing, at least under the summation sign, allow for a Leibniz
rule, associativity, antisymmetry, and Poincare' duality.
\vskip0.5truein
\centerline{\bf References}

\item{1.}A.S.Schwarz, Lett. Math. Phys. 2(1978), 247;
Comm. Math. Phys. 67 (1979), 1.
\item{2.}E.Witten, Comm. Math. Phys. 121 (1989), 351.
\item{3.}E.Witten, Comm. Math. Phys. 117 (1988), 353.
\item{4.}D.Birmingham, M.Blau, M.Rakowski and G.Thompson, `Topological
Field Theory', to be published in Physics Reports, 1992.
\item{5.}Chern-Simons and the Hall effect
\item{6.}A. Fetter, R. Hanna and R. Laughlin, Phys. Rev. B40, 8745 (1989)
\item{7.}J. Fr\"ohlich and P. Marchetti, Comm. Math. Phys. 116 (1988),
127.
\item{8.}M. L\"uscher, Nucl. Phys. B326 (1989), 557.
\item{9.}V.F. M\"uller,  Z.Phys.C47:301-310,1990.
\item{10.}D.Eliezer, G.W.Semenoff and S.S.C.Wu, UBC preprint, 1991.
\item{11.}D.Eliezer and G.W.Semenoff, UBC preprint, 1991.
\item{12.}R. Sorkin, J. Math. Phys. 16 (1975), 2432
\item{13.}T. Honan, Doctoral Thesis, Univ. Maryland, 1985, unpublished
\item{14.}H. Becher and P. Joos, Z. Phys. C15, 343 (1982).
\item{15.}R. Kantor and L. Susskind, `A Lattice Model of Fractional
Statistics', Stanford Preprint, 1990;
D. Eliezer and G.W.Semenoff, Phys. Lett. B266 (1991), 375.

\end